\documentclass{article}

\usepackage{microtype}
\usepackage{graphicx}
\usepackage{subcaption}
\usepackage{booktabs} 
\usepackage{multirow}
\usepackage{xcolor}
\usepackage{colortbl}
\usepackage{tabularx}
\usepackage{tcolorbox}

\usepackage{hyperref}

\usepackage[preprint]{icml2026}

\usepackage{amsmath}
\usepackage{amssymb}
\usepackage{mathtools}
\usepackage{amsthm}

\usepackage[capitalize,noabbrev]{cleveref}

\theoremstyle{plain}

\theoremstyle{definition}

\theoremstyle{remark}

\usepackage[textsize=tiny]{todonotes}

\icmltitlerunning{Bridging Perception and Action}

\begin{document}



\twocolumn[
  \icmltitle{Bridging Perception and Action: \\
    A Lightweight Multimodal Meta-Planner Framework for Robust Earth Observation Agents}


  \icmlsetsymbol{equal}{*}

  \begin{icmlauthorlist}
    \icmlauthor{Jinghui Xu}{sklab,bisie,sigs}
    \icmlauthor{Boyi Shangguan}{sklab,bisie}
    \icmlauthor{Mengke Zhu}{sklab,bisie}
    \icmlauthor{Hao Liu}{sklab,bisie}
    \icmlauthor{Junhuan Jiang}{uts}
    \icmlauthor{Guangjun He}{sklab,bisie}
    \icmlauthor{Pengming Feng}{sklab,bisie}
    \icmlauthor{Shichao Jin}{sklab,bisie}
    \icmlauthor{Bin Liang}{bju}
    \icmlauthor{Yongzhe Chang}{sigs}
    \icmlauthor{Tiantian Zhang}{sigs}
    \icmlauthor{Xueqian Wang}{sigs}
  \end{icmlauthorlist}

  \icmlaffiliation{sklab}{State Key Laboratory of Space Information System and Integrated Application, Beijing 100095, China}
  \icmlaffiliation{bisie}{Beijing Institute of Satellite Information Engineering, Beijing 100095, China}
  \icmlaffiliation{sigs}{Shenzhen International Graduate School, Tsinghua University, Shenzhen, China}
  \icmlaffiliation{uts}{Data Science Institute, University of Technology Sydney, Australia}
  \icmlaffiliation{bju}{School of Electric and Information Engineering, Beijing Jiaotong University, Beijing 100044, China}

  \icmlcorrespondingauthor{Xueqian Wang}{wang.xq@sz.tsinghua.edu.cn}
  \icmlcorrespondingauthor{Tiantian Zhang}{zhang.tt@sz.tsinghua.edu.cn}

  \icmlkeywords{Remote Sensing, Large Multimodal Models, Agents, Meta-Learning, ICML}

  \vskip 0.3in
]



\printAffiliationsAndNotice{}  

\begin{abstract}
Autonomous Earth Observation (EO) agents are transitioning from passive perception to complex, multi-step task execution. However, current architectures that integrate planning and execution within a single model often struggle with combinatorial complexity and reasoning errors in dynamic EO scenarios. To resolve these challenges, we propose the Lightweight Multimodal Meta-Planner (LMMP) framework. LMMP incorporates a dual-awareness mechanism that grounds strategic plans in both multimodal image features and high-level task semantics. Crucially, we introduce a Meta Task Library to inject remote sensing expert knowledge directly into the workflow, which standardizes domain logic and ensures plans are physically feasible. We further implement a two-stage training pipeline, initializing the Meta-Planner via expert-distilled Supervised Fine-Tuning and refining it through Direct Preference Optimization based on execution feedback. Extensive experiments on a dataset derived from EarthBench and ThinkGeo demonstrate that LMMP significantly improves tool-calling accuracy and task success rates. Moreover, the framework exhibits strong ``plug-and-play'' versatility, consistently enhancing the performance of diverse executor backbones across previously unseen EO missions. Code is available at: \url{https://anonymous.4open.science/r/anonymous-EO-MetaPlanner-FCBD}.
\end{abstract}

\section{Introduction}
\label{sec:intro}    

The rapid proliferation of remote sensing satellites and the accelerating pace of data acquisition are driving a profound paradigm shift in Earth Observation (EO): from static, task-specific recognition toward dynamic, agentic intelligence~\cite{kuckreja2024geochat, zhang2024earthgpt, zhan2025skyeyegpt}. This evolution has progressed through several distinct stages. In the early 2010s, the field relies on specialized models tailored for supervised tasks like land-cover classification and object detection~\cite{xia2017aid, cheng2017remote, sumbul2019bigearthnet, liu2024remoteclip, wang2025sarclip}. While effective within narrow domains, these models lacked the generalization capabilities required for diverse sensors and complex scenarios. Around 2020, the emergence of Remote Sensing Foundation Models (RSFMs) marked a pivotal turn~\cite{sun2022ringmo, yan2023ringmo, guo2024skysense, zhan2025skyeyegpt, diao2025ringmo, zhang2025skysense, wu2025semantic, hu2025rsgpt}. Leveraging large-scale self-supervised pre-training, RSFMs learned versatile, general-purpose EO representations. This progress was further catalyzed by Remote Sensing Multimodal Large Language Models (RS-MLLM)~\cite{kuckreja2024geochat, muhtar2024lhrs, zhang2024earthgpt, soni2025earthdial}, which revolutionized EO data interaction by aligning multimodal encoders with the reasoning capabilities of LLMs, enabling a leap from passive perception to interactive, query-driven interpretation. Today, we stand at the dawn of autonomous EO agents—systems designed to reason proactively, utilize tools, and execute complex, multi-step missions with minimal human intervention~\cite{guo2024remote, xu2024rs, hu2025ringmo, wei2025geotool}.

Despite this transformative potential, a critical robustness gap impedes the transition of EO agents from prototypes to reliable, mission-critical systems. A core limitation lies in current agent architectures that integrate planning and execution within a single model, where a general-purpose LLM is tasked with both high-level reasoning and precise tool execution~\cite{singh2024geollm, guo2024remote}. This integrated design creates a significant bottleneck: when confronted with extensive, specialized toolkits, the agent faces combinatorial ``action space explosion'' and cognitive overload~\cite{qin2023toolllm}. Consequently, these systems frequently suffer from reasoning failures, such as hallucinated steps or execution errors, particularly in long-horizon tasks requiring precise sequential logic. Moreover, general-purpose models often lack an intrinsic understanding of domain-specific constraints, leading to plans that are syntactically plausible yet semantically or physically infeasible~\cite{singh2024geollm}. This perception-action misalignment critically undermines the operational reliability of autonomous EO Agents.

To address these challenges, we propose the Lightweight Multimodal Meta-Planner (LMMP) framework. Our core design philosophy is Cognitive Decoupling: instead of forcing a single model to handle both high-level strategy and low-level execution, we introduce a specialized meta-planning layer. This layer functions as a ``domain expert'' that connects raw perception with complex action~\cite{hasan2025mapagent, xiong2025mpo}. The LMMP incorporates a dual-awareness mechanism that natively integrates the perceptual outputs of an RS-MLLM into the planning loop. This integration allows the Meta-Planner to jointly reason over visual context and high-level task semantics, ensuring that planning is a visually-grounded process. Crucially, the framework acts as a ``Plug-and-Play'' cognitive module that navigates the vertical domain logic for generalist executors. We further introduce a Meta Task Library that injects domain expert knowledge. Guided by the planner's strategic intent, this library dynamically prunes the tool search space, delivering only context-relevant tools to the executor. For functional reliability, we implement a two-stage training pipeline, initializing the Meta-Planner via expert-distilled Supervised Fine-Tuning (SFT)~\cite{muhtar2024lhrs, li2025lhrs} and refining it through execution-feedback-driven Direct Preference Optimization (DPO)~\cite{rafailov2023direct, qian2025toolrl}. This integrated pipeline ensures that the LMMP generates action trajectories that are logically coherent and robustly executable.

The principal contributions of this work are three-fold:
\begin{itemize}
    \item We propose the LMMP framework that leverages RS-MLLM perception to empower a dual-awareness Meta-Planner. By establishing a dedicated layer for strategic reasoning, the framework mitigates cognitive overload and transforms EO data perception into grounded, context-aware action sequences.
    \item We develop a two-stage training pipeline for the Meta-Planner. This process combines SFT based on expert logic distillation with DPO driven by fine-grained execution feedback, ensuring generated plans are both logically sound and functionally robust.
    \item We demonstrate the ``plug-and-play'' versatility of our framework through extensive evaluations on EarthBench~\cite{feng2025earth} and ThinkGeo~\cite{shabbir2025thinkgeo}. Our results show that the specialized Meta-Planner significantly enhances tool-calling accuracy across diverse executor backbones and unseen missions, proving effectively as a scalable cognitive patch for generalist models.
\end{itemize}

\begin{figure}[ht]
    \vskip -0.1in
    \centering  
    \includegraphics[width=\columnwidth]{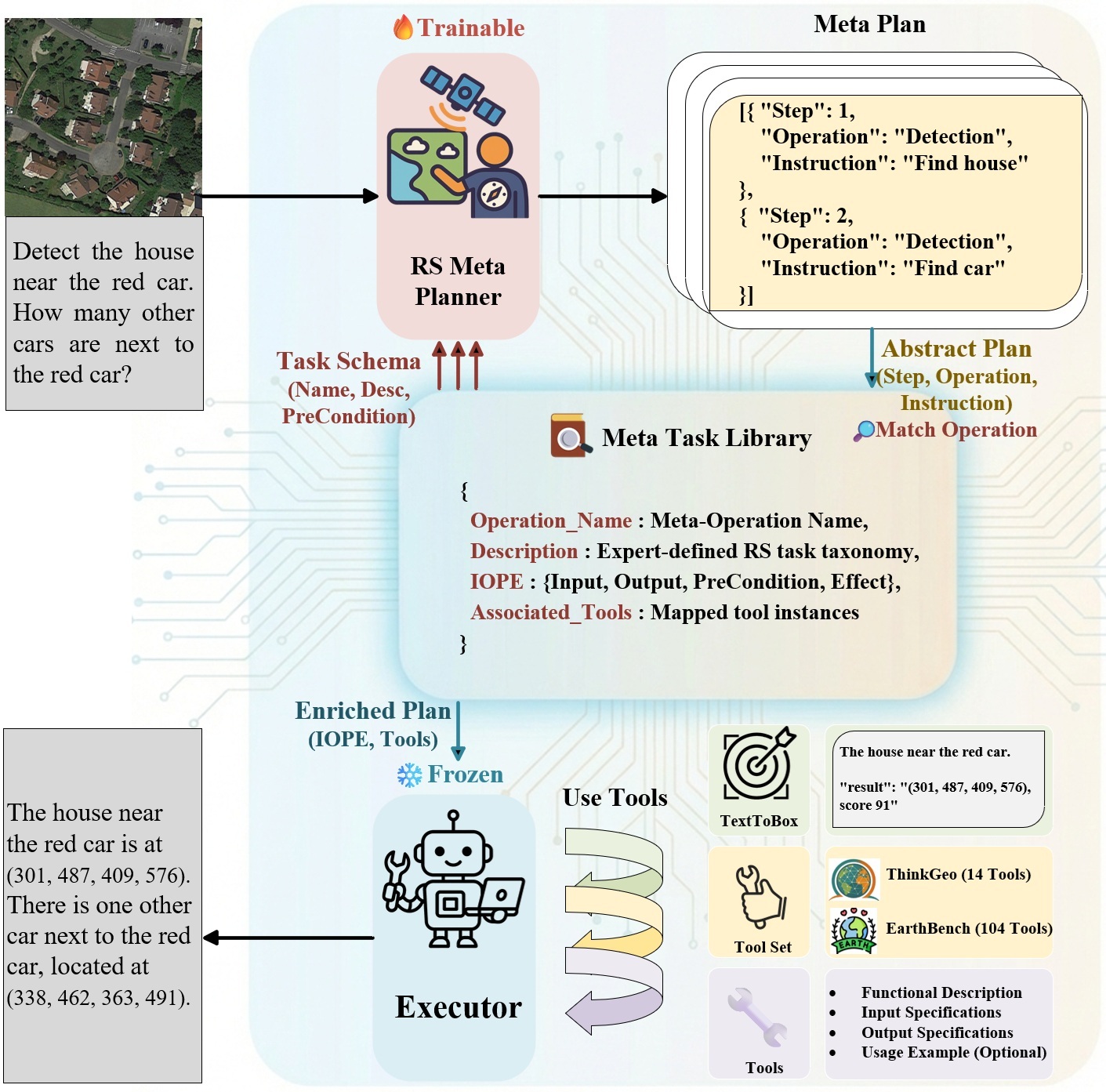}
    \caption{Overview of the LMMP framework. The trainable RS Meta-Planner generates a structured Meta Plan based on multimodal input. This plan is then enriched with domain-specific logic from the Meta Task Library, allowing a frozen Executor to call relevant tools and generate the final response.}
    \label{fig:framework}
    \vskip -0.22in 
\end{figure}

\section{Methodology}
\label{sec:methodology}

This section presents the Lightweight Multimodal Meta-Planner framework. We design this architecture to resolve the reasoning limitations and cognitive overload often observed in agents that integrate planning and execution within a single model. The essence of our approach lies in Semantic Abstraction: by elevating decision-making from atomic tool selection to high-level strategic planning, we drastically reduce the complexity of the action space. To achieve robust performance in complex EO missions, our approach relies on three complementary mechanisms. Firstly, we leverage the perception capabilities of an RS-MLLM to ground planning steps in the specific visual context. Secondly, we incorporate a Meta Task Library to inject explicit remote sensing expert knowledge directly into the workflow, abstracting over 100 tools into coherent operational categories. Thirdly, we employ a two-stage training pipeline that allows the Meta-Planner to internalize domain logic and tool usage patterns. Finally, we introduce the process-outcome assessment used for evaluation.

\begin{figure*}[t!]
    \centering
    \includegraphics[width=\textwidth]{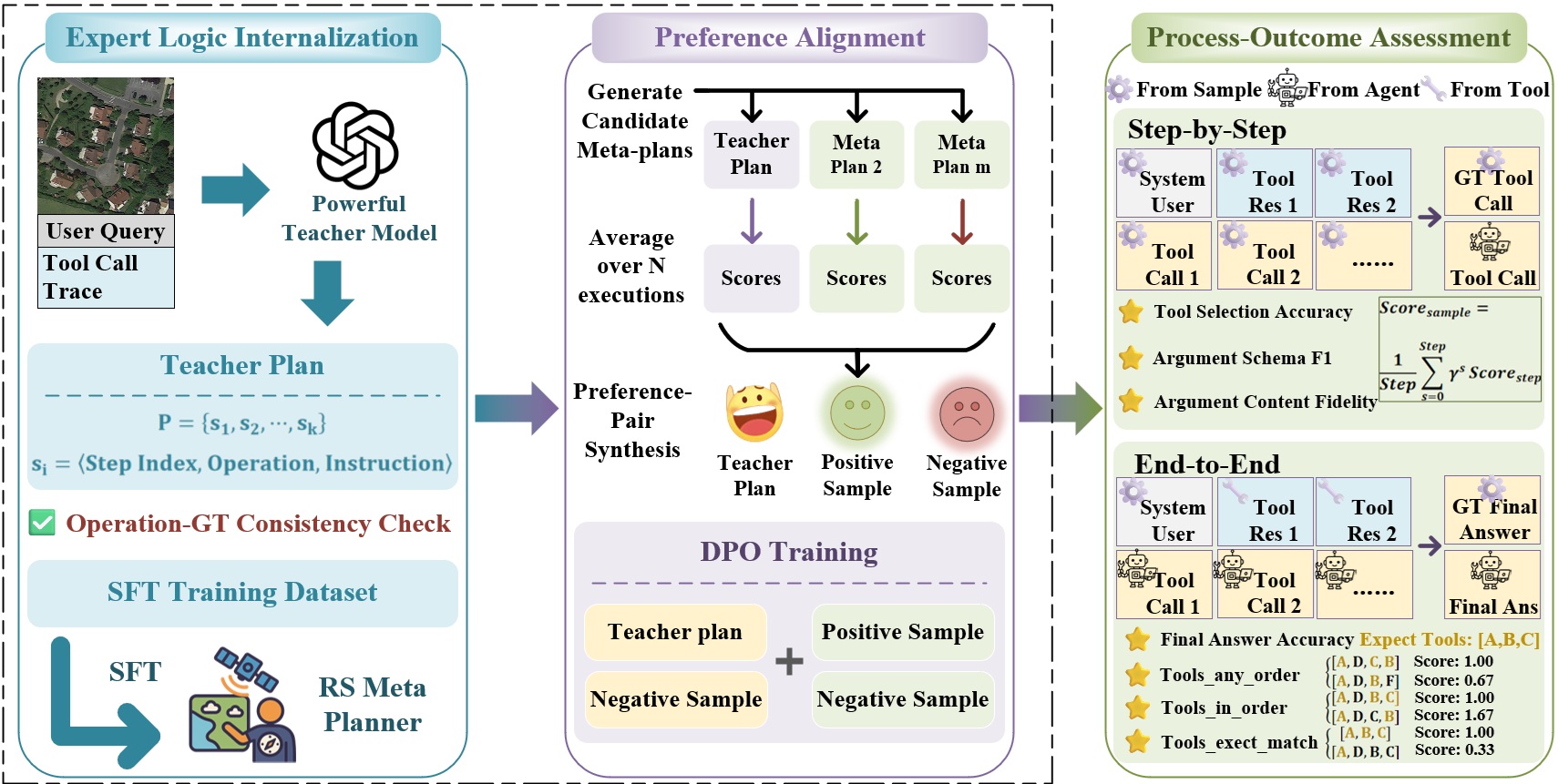}
    \caption{The training and evaluation pipeline. The two-stage evolutionary training consists of  Expert Logic Internalization (Stage 1) and Preference Alignment (Stage 2). The Process-Outcome Assessment shows how the Meta-Planner is evaluated through step-by-step and end-to-end metrics.}
    \label{fig:training_pipeline}
\end{figure*}

\subsection{LMMP Framework}
\label{sec:framework}

To resolve the misalignment between abstract user intent and concrete tool usage, the LMMP framework strictly separates high-level reasoning from precise execution. As illustrated in Figure~\ref{fig:framework}, the system processes multimodal queries through three specialized layers.

\subsubsection{RS Meta-Planner}
The \textbf{RS Meta-Planner} ($\pi_{P}$) serves as the reasoning core of the system. Unlike conventional agents~\cite{xu2024rs} that map queries directly to API calls, it focuses on synthesizing a coarse-grained meta plan $P = \{s_1, s_2, ..., s_k\}$. We employ a Dual-Awareness Mechanism that synergizes task semantics awareness with spatio-spectral awareness. This design ensures that the Meta-Planner interprets high-level user intent from textual queries while simultaneously extracting spatial constraints from the imagery. Such visual grounding is essential because remote sensing data differs fundamentally from natural photography through its intricate spatial details and drastic scale variations. By accounting for this unique visual context, the Meta-Planner generates instructions that are not only logically aligned with the user goal but also tailored to the specific scene configuration. Each meta step $s_i$ consists of three elements: (1) Step Index to mark the temporal order; (2) Operation which is a semantic abstraction representing a functional task category; and (3) Instruction containing natural language guidance to advise the executor.

\subsubsection{Meta Task Library}
The \textbf{Meta Task Library} serves as an external repository of remote sensing domain expertise. While the Meta-Planner provides the strategic direction, this library injects the necessary expert knowledge required for execution. It stores standardized task definitions including ``IOPE'' specifications (Input, Output, Pre-condition, Effect) and maps abstract operations to specific tool configurations. During the inference process, the library aligns the planner's output with these expert definitions. This enrichment step ensures that the final plan passed to the executor is not only logically sound but also technically precise. As a secondary benefit, this mechanism automatically identifies the relevant tool subsets for each operation, which effectively reduces the search space and noise for the executor.

\subsubsection{Executor}
The \textbf{Executor} ($\pi_{E}$) operates in a standard ReAct loop to perform concrete actions. Instead of facing the full tool space, it receives an enriched meta plan containing injected domain knowledge and a focused list of tools. The executor is strictly responsible for translating the semantic instruction into precise executable code or API calls. It also reports execution observations back to the system to maintain global state consistency.

\subsection{Two-Stage Training Pipeline}
\label{sec:training}

To enable the Meta-Planner to internalize domain expertise and adapt to the physical constraints of tool execution, we implement a two-stage training pipeline. As illustrated in Figure~\ref{fig:training_pipeline}, this process follows a clear hierarchy: firstly establishing foundational reasoning capabilities via Expert Logic Internalization, then refining the policy through Preference Alignment based on execution feedback.

\subsubsection{Stage 1: Expert Logic Internalization}
The objective of this stage is to train the LMMP to generate valid meta plans by mimicking expert behavior. We construct a dataset of high-quality meta plans using knowledge distillation from a teacher model.

For each training sample, the input consists of the user query $u$ and the resized RS imagery $I$. The teacher model processes the complete ground-truth execution history, including tool call traces and observations, to synthesize a structured Teacher Plan $p = \{s_1, ..., s_k\}$.

To ensure the reliability of this distilled knowledge, we enforce a strict quality control mechanism. Each synthesized plan is verified against the Meta Task Library (Sec.~\ref{sec:framework}). We retain only those meta plans where the operation sequence strictly adheres to the library's logical constraints. The LMMP is then optimized on this verified dataset $D_{sft}$ via the standard auto-regressive objective:
\begin{equation}
    \mathcal{L}_{SFT} = -\mathbb{E}_{(u,I,p)\sim D_{sft}}[\log \pi_{\theta}(p|u, I)]
\end{equation}
This stage effectively initializes the LMMP's Dual-awareness Mechanism, enabling the model to correlate learned geospatial planning logic with multimodal image features.

\subsubsection{Stage 2: Preference Alignment}
While SFT establishes the foundational planning logic, the model may still produce plans that are logically plausible but fail during actual execution. To resolve this discrepancy, we employ preference optimization driven by execution outcomes.

\paragraph{Robust Preference Construction:}
We employ the SFT-initialized model to sample multiple candidate plans $\{p_1, ..., p_M\}$ for a given input. Given the nondeterministic nature of tool execution in dynamic environments, we execute each candidate several times to obtain a reliable performance distribution. We evaluate these trajectories using a Step-aware Discounted Reward, where the score at step $t$ is weighted by $\gamma^{t-1}$ ($\gamma=0.9$; see Appendix~\ref{sec:appendix_gamma} for the hyperparameter sensitivity analysis) to prioritize early-stage correctness, which serves as the precondition for subsequent success.

To construct the preference dataset $\mathcal{D}_{\text{DPO}}$, we select pairs $(p_w, p_l)$ based on statistical validity rather than single-run variance. A pair is formed only if the score distribution of $p_w$ is superior to $p_l$ with statistical significance. This ensures that the optimization signal reflects functional improvement.

\paragraph{Hybrid Optimization:}
To balance exploration and stability, we utilize a hybrid preference dataset containing: (1) Self-Generated Pairs sampled from the current policy to encourage exploration, and (2) Teacher-Augmented Pairs, where the validated ground-truth plan acts as a strong positive anchor ($p_w$) against weaker sampled plans. This hybrid design ensures the optimized LMMP generalizes well to unseen EO tasks while retaining reliable reasoning logic. The model is optimized using the DPO objective:
\begin{equation}
    \mathcal{L}_{\text{DPO}} = -\mathbb{E}_{(x, p_w, p_l) \sim \mathcal{D}} \left[ \log \sigma \left( r_\theta(x, p_w) - r_\theta(x, p_l) \right) \right]
\end{equation}
where $r_\theta(x, y) = \beta \log \frac{\pi_\theta(y|x)}{\pi_{\text{ref}}(y|x)}$ represents the implicit reward ratio, $\beta$ is the temperature parameter controlling preference strength.

\subsection{Process-Outcome Assessment}
\label{sec:evaluation_metrics}

To rigorously assess both the logical validity of the action trajectory and the functional correctness of the final execution, we establish a set of Process-Outcome Assessment metrics. As illustrated in the green panel of the Figure~\ref{fig:training_pipeline}, these metrics are classified into two complementary dimensions: Step-by-Step metrics for local reasoning verification, and End-to-End metrics for global mission success.

\subsubsection{Step-by-Step Process Evaluation}
Drawing on established methodologies from ToolRL~\cite{qian2025toolrl} and ThinkGeo~\cite{shabbir2025thinkgeo}, this phase evaluates the agent's immediate reasoning fidelity under a \textit{teacher-forcing} regime. To capture the cascading impact of early-stage decisions, we verify the correctness of the predicted action $\hat{a}_t$ at each time step $t$ given the ground-truth history. We aggregate these step-wise scores using a discounted reward mechanism to compute the sample-level process score $S_{sample}$:
\begin{equation}
    S_{sample} = \frac{1}{\sum_{t=1}^T \gamma^{t-1}} \sum_{t=1}^T \gamma^{t-1} \cdot \mathcal{M}_{step}(t)
\end{equation}
where $T$ denotes the trajectory length, and $\gamma=0.9$ is the discount factor emphasizing proximal reasoning steps. The step-wise metric $\mathcal{M}_{step}$ is defined hierarchically:

\begin{enumerate}
    \item \textbf{Tool Selection Accuracy (TSA):} A strict binary measure validating if the predicted tool name $\hat{a}_t$ matches the ground truth $a_t$ exactly.
    \item \textbf{Argument Schema F1 (ASF1):} Conditioned on correct tool selection, this metric assesses whether the set of predicted parameter keys $K_{pred}$ aligns with the ground truth keys $K_{gt}$, computed via the F1-score.
    \item \textbf{Argument Content Fidelity (ACF):} Conditioned on a correct schema, this measures the semantic consistency of the parameter values. We employ a data-type-aware scoring function $\phi(v_{pred}, v_{gt})$:
    \begin{equation}
    \phi = 
    \begin{cases} 
      \text{F1-Score}(v_{p}, v_{g}) & \text{if type is List/Tuple} \\
      \text{ROUGE-L}(v_{p}, v_{g}) & \text{if type is String} \\
      \mathbb{I}(v_{p} = v_{g}) & \text{otherwise}
    \end{cases}
    \end{equation}
\end{enumerate}

\subsubsection{End-to-End Outcome Evaluation}
Complementary to process verification, we employ outcome-oriented metrics proposed by Earth-Agent~\cite{feng2025earth} to assess the global success of the mission.

\paragraph{Final Answer Accuracy (FAA) via Semantic Scoring:}
This metric evaluates the correctness of the agent's final response against the Ground Truth (GT). While standard multiple-choice tasks (e.g., EarthBench) utilize strict binary accuracy, the heterogeneous nature of open-ended RS queries (e.g., ThinkGeo) renders exact string matching ineffective. 

To address this, we establish a Task-Adaptive Semantic Scoring Protocol designed by domain experts. As detailed in Table~\ref{tab:judge_protocol}, this protocol categorizes queries into distinct types such as MCQ, Numerical, Boolean, and Description. For each type, the judge applies rigorous domain constraints to calculate a functional correctness score $S \in [0, 1]$. This approach prioritizes the recall of Key Information Points (KIPs), ensuring valid but paraphrased answers are credited while penalizing specific hallucinations.

\paragraph{Tool Sequence Alignment:}
To evaluate the macro-level planning logic independent of parameter details, we compare the sequence of predicted tool names $\mathcal{T}_{pred}$ against the ground-truth expert sequence $\mathcal{T}_{gt}$:
\begin{itemize}
    \item \textbf{Tools Any Order (TAO):} Measures the recall of required tools regardless of execution order: $|\mathcal{T}_{pred} \cap \mathcal{T}_{gt}| / |\mathcal{T}_{gt}|$.
    \item \textbf{Tools In Order (TIO):} Evaluates the correct ordering of tools while tolerating non-critical gaps. It is computed using the Longest Common Subsequence: $|LCS(\mathcal{T}_{pred}, \mathcal{T}_{gt})| / |\mathcal{T}_{gt}|$.
    \item \textbf{Tools Exact Match (TEM):} The strictest metric, requiring the execution trace to perfectly mirror the expert demonstration from the initial step. It is derived from the Longest Common Prefix: $|LCP(\mathcal{T}_{pred}, \mathcal{T}_{gt})| / |\mathcal{T}_{gt}|$.
\end{itemize}

\begin{table}[t]
\caption{The LLM-as-a-Judge evaluation protocol. We define specific scoring criteria and tolerance thresholds for distinct geospatial task types to ensure consistent and rigorous automated evaluation.}
\label{tab:judge_protocol}
\begin{center}
\begin{small}
\renewcommand{\arraystretch}{1.25} 
\resizebox{\columnwidth}{!}{ 
\begin{tabular}{l p{0.3\columnwidth} p{0.6\columnwidth}}
\toprule
\textbf{Task Category} & \textbf{Scoring Metric} & \textbf{Evaluation Criteria \& Tolerance} \\
\midrule
\textbf{Multiple Choice} & Strict Binary Match & Exact match of the option identifier (A/B/C/D). Correct reasoning with a wrong final selection yields a score of 0. \\
\midrule
\textbf{Numerical} & Value Recall Rate & Ratio of correctly extracted quantities. Integers/Counts must be exact; Float values allow a $\pm5\%$ deviation tolerance. \\
\midrule
\textbf{Boolean} & Semantic Alignment & Semantic verification of affirmative or negative intent (e.g., matching ``Yes'' with ``True''), ignoring syntactic variations. \\
\midrule
\textbf{Description} & Key Info Point (KIP) Recall & Score computed as $\frac{\text{Matched KIPs}}{\text{Total GT KIPs}}$. The judge extracts core entities/states (Nouns/Verbs) while ignoring generic adjectives and non-contradictory extra information. \\
\bottomrule
\end{tabular}
}
\end{small}
\end{center}
\vskip -0.1in
\end{table}

\begin{table}[t!]
\centering
\caption{Statistics of the datasets used in our experiments. ``Unseen Tools'' refers to tools present in the test set that are not accessible during training. ``Avg. Steps'' denotes the average number of tool calls per sample.}
\label{tab:dataset_stats}
\resizebox{\columnwidth}{!}{%
\begin{tabular}{l|c|c|c|c|c}
\toprule
\textbf{Dataset} & \textbf{Split} & \textbf{Samples} & \textbf{Total Tools} & \textbf{Unseen Tools} & \textbf{Avg. Steps} \\
\midrule
\multirow{2}{*}{EarthBench} & Train & 200 & 45 & - & 5.39 \\
 & Test & 46 & 76 & 31 & 5.11 \\
\midrule
\multirow{2}{*}{ThinkGeo} & Train & 330 & 14 & - & 3.49 \\
 & Test & 57 & 14 & 0 & 3.12 \\
\midrule
GeoScenario-116 & Test & 116 & 9 & 9 & 3.06 \\
\bottomrule
\end{tabular}%
}
\end{table}

\section{Experiments}
\label{sec:experiments}

\definecolor{rowgray}{gray}{0.95} 
\definecolor{upcolor}{RGB}{200, 0, 0} 
\definecolor{downcolor}{RGB}{0, 128, 0} 
\newcommand{\up}{\textcolor{upcolor}{$\uparrow$}}
\newcommand{\down}{\textcolor{downcolor}{$\downarrow$}}

\begin{table*}[t]
\caption{Detailed performance comparison across EarthBench, ThinkGeo, and GeoScenario-116. Metrics are grouped into \textbf{Process Metrics} (TSA, ASF1, ACF) and \textbf{Outcome Metrics} (FAA, TAO, TIO, TEM). \textbf{Bold} indicates the best result, and \underline{underlined} indicates the second best. Columns with a gray background denote our approach.}
\label{tab:detailed_results}
\centering
\small
\setlength{\tabcolsep}{0pt} 
\renewcommand{\arraystretch}{1.1} 

\begin{tabular*}{\textwidth}{@{\extracolsep{\fill}} l cc cc cc c cc cc cc cc @{}}
\toprule
\multirow{2.5}{*}{\textbf{Model}} 
& \multicolumn{2}{c}{\textbf{TSA}} & \multicolumn{2}{c}{\textbf{ASF1}} & \multicolumn{2}{c}{\textbf{ACF}} 
& & \multicolumn{2}{c}{\textbf{FAA}} & \multicolumn{2}{c}{\textbf{TAO}} & \multicolumn{2}{c}{\textbf{TIO}} & \multicolumn{2}{c}{\textbf{TEM}} \\
\cmidrule{2-7} \cmidrule{9-16}
& Base & \cellcolor{rowgray}Meta & Base & \cellcolor{rowgray}Meta & Base & \cellcolor{rowgray}Meta 
& & Base & \cellcolor{rowgray}Meta & Base & \cellcolor{rowgray}Meta & Base & \cellcolor{rowgray}Meta & Base & \cellcolor{rowgray}Meta \\
\midrule

\multicolumn{16}{l}{\textit{\textbf{EarthBench} (104 Tools)}} \\
\midrule
Qwen3-8B    
& 0.399 & \cellcolor{rowgray} 0.629 \up & 0.369 & \cellcolor{rowgray}0.592\up & 0.279 & \cellcolor{rowgray}\underline{0.530}\up 
&& 0.065 & \cellcolor{rowgray}0.174\up & 0.448 & \cellcolor{rowgray}0.531\up & 0.358 & \cellcolor{rowgray}0.440\up & 0.034 & \cellcolor{rowgray}0.261\up \\

Qwen3-235B  
& 0.537 & \cellcolor{rowgray}\textbf{0.698}\up & 0.479 & \cellcolor{rowgray}\textbf{0.655}\up & 0.421 & \cellcolor{rowgray}\textbf{0.604}\up 
&& 0.261 & \cellcolor{rowgray}0.326\up & 0.618 & \cellcolor{rowgray}0.736\up & 0.458 & \cellcolor{rowgray}0.612\up & 0.183 & \cellcolor{rowgray}0.429\up \\

Qwen2.5-72B 
& 0.489 & \cellcolor{rowgray}\underline{0.643}\up & 0.447 & \cellcolor{rowgray}\underline{0.621}\up & 0.347 & \cellcolor{rowgray}0.523\up 
&& 0.370 & \cellcolor{rowgray}0.355\down & 0.668 & \cellcolor{rowgray}0.715\up & 0.518 & \cellcolor{rowgray}0.553\up & 0.066 & \cellcolor{rowgray}0.308\up \\

GPT-5       
& 0.541 & \cellcolor{rowgray}0.511\down & 0.505 & \cellcolor{rowgray}0.488\down & 0.410 & \cellcolor{rowgray}0.432\up 
&& \underline{0.565} & \cellcolor{rowgray}\textbf{0.609}\up & \underline{0.770} & \cellcolor{rowgray}\textbf{0.784}\up & \underline{0.649} & \cellcolor{rowgray}\textbf{0.663}\up & \textbf{0.526} & \cellcolor{rowgray}\underline{0.500}\down \\

\midrule
\multicolumn{16}{l}{\textit{\textbf{ThinkGeo} (14 Tools)}} \\
\midrule
Qwen3-8B    
& 0.502 & \cellcolor{rowgray}\underline{0.743}\up & 0.501 & \cellcolor{rowgray}\underline{0.741}\up & 0.348 & \cellcolor{rowgray}\underline{0.519}\up 
&& 0.199 & \cellcolor{rowgray}0.246\up & 0.769 & \cellcolor{rowgray}0.769~ & 0.606 & \cellcolor{rowgray}0.659\up & 0.148 & \cellcolor{rowgray}0.506\up \\

Qwen3-235B  
& 0.483 & \cellcolor{rowgray}\textbf{0.777}\up & 0.455 & \cellcolor{rowgray}\textbf{0.743}\up & 0.392 & \cellcolor{rowgray}\textbf{0.550}\up 
&& 0.290 & \cellcolor{rowgray}0.307\up & \underline{0.845} & \cellcolor{rowgray}\textbf{0.886}\up & 0.691 & \cellcolor{rowgray}\textbf{0.767}\up & 0.348 & \cellcolor{rowgray}\textbf{0.625}\up \\

Qwen2.5-72B 
& 0.511 & \cellcolor{rowgray}0.720\up & 0.511 & \cellcolor{rowgray}0.720\up & 0.344 & \cellcolor{rowgray}0.497\up 
&& 0.263 & \cellcolor{rowgray}0.219\down & 0.718 & \cellcolor{rowgray}0.727\up & 0.592 & \cellcolor{rowgray}0.625\up & 0.080 & \cellcolor{rowgray}0.395\up \\

GPT-5       
& 0.484 & \cellcolor{rowgray}0.529\up & 0.484 & \cellcolor{rowgray}0.528\up & 0.293 & \cellcolor{rowgray}0.259\down 
&& \textbf{0.535} & \cellcolor{rowgray}\underline{0.400}\down & 0.709 & \cellcolor{rowgray}0.841\up & 0.571 & \cellcolor{rowgray}\underline{0.731}\up & 0.391 & \cellcolor{rowgray}\underline{0.509}\up \\

\midrule
\multicolumn{16}{l}{\textit{\textbf{GeoScenario-116} (9 Tools)}} \\
\midrule
Qwen3-8B    
& 0.348 & \cellcolor{rowgray}\textbf{0.738}\up & 0.347 & \cellcolor{rowgray}\textbf{0.735}\up & 0.265 & \cellcolor{rowgray}\underline{0.531}\up 
&& 0.383 & \cellcolor{rowgray}0.302\down & 0.560 & \cellcolor{rowgray}0.879\up & 0.487 & \cellcolor{rowgray}0.835\up & 0.214 & \cellcolor{rowgray}0.617\up \\

Qwen3-235B  
& 0.461 & \cellcolor{rowgray}\underline{0.685}\up & 0.459 & \cellcolor{rowgray}\underline{0.685}\up & 0.379 & \cellcolor{rowgray}\textbf{0.547}\up 
&& \textbf{0.399} & \cellcolor{rowgray}0.356\down & 0.722 & \cellcolor{rowgray}\underline{0.966}\up & 0.690 & \cellcolor{rowgray}\underline{0.917}\up & \underline{0.486} & \cellcolor{rowgray}\textbf{0.760}\up \\

Qwen2.5-72B 
& 0.416 & \cellcolor{rowgray}0.668\up & 0.404 & \cellcolor{rowgray}0.663\up & 0.315 & \cellcolor{rowgray}0.491\up 
&& \underline{0.391} & \cellcolor{rowgray}0.365\down & 0.655 & \cellcolor{rowgray}\textbf{0.970}\up & 0.564 & \cellcolor{rowgray}\textbf{0.923}\up & 0.204 & \cellcolor{rowgray}0.522\up \\

\bottomrule
\end{tabular*}
\begin{flushleft}
\scriptsize 
\textbf{Note:} \textbf{TSA}: Tool Selection Acc; \textbf{ASF1}: Argument Schema F1; \textbf{ACF}: Argument Content Fidelity; \textbf{FAA}: Final Answer Acc; \textbf{TAO}: Tools Any Order; \textbf{TIO}: Tools In Order; \textbf{TEM}: Tools Exact Match.
$\uparrow$ (red) denotes improvement, $\downarrow$ (green) denotes decline compared to baseline.
\end{flushleft}
\end{table*}

\subsection{Experimental Settings}

\subsubsection{Datasets}
We evaluate the LMMP framework on EarthBench~\cite{feng2025earth} and ThinkGeo~\cite{shabbir2025thinkgeo}, which currently represent the state-of-the-art benchmarks dedicated to remote sensing agents. EarthBench focuses on diverse tool usage across perception and physical inversion tasks, whereas ThinkGeo emphasizes spatial reasoning and interpretation. Additionally, we introduce a custom out-of-domain dataset, GeoScenario-116, to assess model robustness in specialized environments such as airports and harbors that differ from the training distribution. Table~\ref{tab:dataset_stats} presents the statistical details for each dataset. Appendix~\ref{app:datasets} provides specific descriptions of data curation and task definitions.

\subsubsection{Implementation Details}
The RS Meta-Planner uses Qwen3-VL-8B-Instruct as its backbone. To ensure training stability, we apply Low-Rank Adaptation (LoRA) configured with a rank of 32 and an alpha of 16. The training procedure adheres to the two-stage pipeline described in Section~\ref{sec:training}.

In the expert logic internalization phase, the model is fine-tuned on a mixture of EarthBench and ThinkGeo training sets for 2 epochs. We use the AdamW optimizer with an initial learning rate of $1 \times 10^{-4}$. The subsequent preference alignment phase optimizes the Meta-Planner using a hybrid dataset of 512 samples, consisting of equal parts self-generated and teacher-augmented pairs. This stage spans 1 epoch with a learning rate of $1 \times 10^{-4}$ and a KL penalty coefficient $\beta$ set to 0.1.

We evaluate open-source models on a local offline server cluster to maintain environmental consistency. For proprietary models, we adopt a two-step pipeline where valid meta-plans are generated offline and executed via API calls. All experiments run on a computing server with 8 NVIDIA A100-80G GPUs connected by NVLINK.

\subsection{Effectiveness of Hierarchical Meta-Planning}
\label{sec:main_results_analysis}

Table~\ref{tab:detailed_results} presents the performance comparison across EarthBench, ThinkGeo, and GeoScenario-116. The integration of the LMMP framework consistently improves agent performance across all datasets and backbone models. On the complex EarthBench dataset, the Qwen3-8B agent achieves a TSA increase from 39.9\% to 62.9\%. This performance rivals the much larger Qwen2.5-72B baseline (48.9\%), suggesting that the specialized Meta-Planner effectively compensates for the reasoning limitations of smaller models. Regarding outcome metrics, the framework significantly enhances temporal consistency. For instance, on ThinkGeo, the Qwen3-8B baseline achieves a TEM of only 0.148, while the Meta-Planner elevates to 0.506. Furthermore, our approach refines even state-of-the-art generalist models, boosting the FAA of GPT-5 on EarthBench from 0.565 to 0.609.

Our method demonstrates robust generalization in out-of-distribution scenarios. On GeoScenario-116, which contains unseen airport and harbor scenes, the Qwen3-8B agent doubles its TSA from 0.348 to 0.738. This improvement is largely attributable to the Meta Task Library, which utilizes expert knowledge to prune the search space to relevant subsets. The consistent gains in the zero-shot tool subset of EarthBench further indicate that the Meta-Planner abstracts the logic of tool usage rather than merely memorizing tool names.

\subsection{Robustness Across Task Complexity}

\begin{figure}[t]
    \vskip 0.2in
    \begin{center}
    \includegraphics[width=1.0\columnwidth]{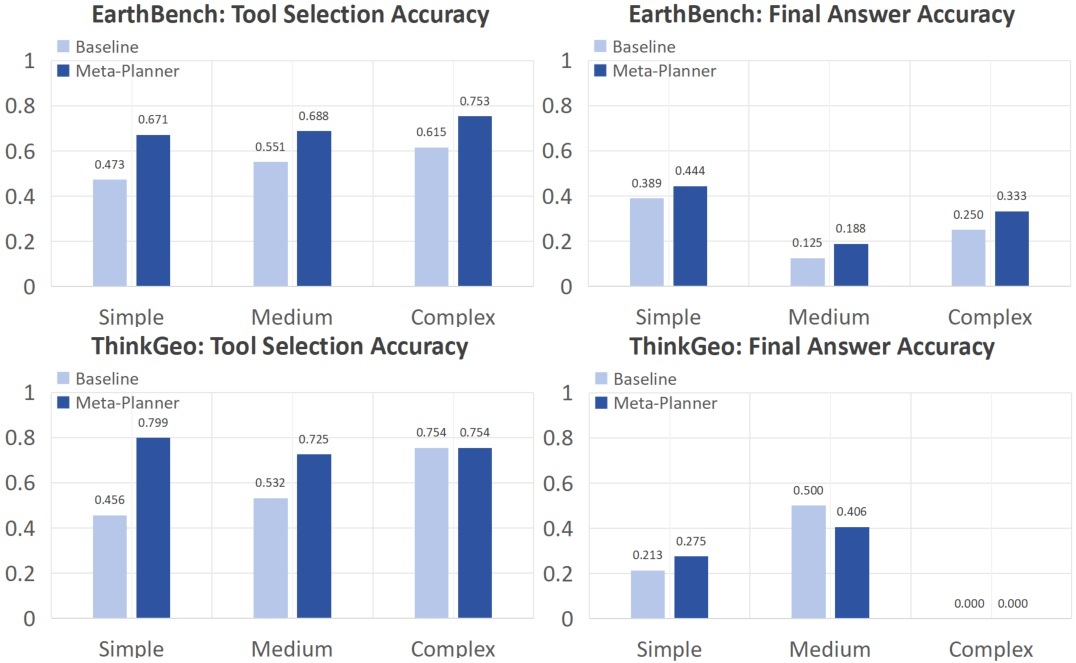}
    \caption{Performance comparison across task complexity levels on EarthBench and ThinkGeo. The complexity levels are defined by the number of steps in the ground truth trajectory: Simple (2-3 steps), Medium (4-6 steps), and Complex ($\ge$7 steps).}
    \label{fig:complexity_analysis}
    \end{center}
    \vskip -0.2in
\end{figure}

To assess performance as reasoning chains lengthen, we categorized test samples into three complexity levels based on the number of steps required: Simple (2-3 steps), Medium (4-6 steps), and Complex (7 steps and above). The EarthBench dataset is divided into 18, 16, and 12 samples by three categories, respectively, while the ThinkGeo dataset is divided into 40, 16, and 1 samples. Figure \ref{fig:complexity_analysis} presents the comparison for TSA and FAA.

Regarding TSA, the Meta-Planner consistently outperforms the baseline across all levels. Interestingly, accuracy on EarthBench increases with task complexity rather than decreasing. For instance, the baseline TSA rises from 0.473 in Simple tasks to 0.615 in Complex tasks, while the Meta-Planner further elevates this to 0.753. This trend occurs because the step-by-step evaluation uses ground-truth history; longer trajectories provide richer context, reducing ambiguity for subsequent tool prediction. The Meta-Planner effectively leverages this context to maintain a significant lead.

The FAA results exhibit higher variance due to the strict requirements of end-to-end execution. While the Meta-Planner generally yields higher success rates, we observe anomalies such as the Medium category in ThinkGeo, where the baseline slightly surpasses the Meta-Planner (0.50 vs 0.41). This discrepancy stems from two primary factors. First, the limited sample size in specific categories introduces statistical noise. Second, execution failures often originate from the inherent instability of the tools rather than planning errors. Current remote sensing tools for object counting or text-to-box generation, may produce incorrect outputs even when invoked correctly. Consequently, a meta plan which are logically sound does not always guarantee a correct final answer if the underlying tool fails.

\subsection{Ablation Study of Training Strategies}

We conduct an ablation study on the EarthBench dataset using the Qwen3-235B executor to isolate the contributions of our architectural design and training stages (Table \ref{tab:ablation_main}). To validate that the Meta Task Library provides expert knowledge injection rather than simple information filtering, we introduce a Tool-RAG baseline. This baseline retrieves top-k tools based on semantic similarity but lacks the hierarchical planning structure.

As shown in Table \ref{tab:ablation_main}, the Base configuration represents the standalone executor operating without the Meta-planner. The executor with Zero-Shot Meta-Planner significantly outperforms the Tool-RAG baseline (TSA 0.626 vs. 0.567). This performance gap confirms that our framework injects structural expert logic into the workflow, which is more effective than merely pruning the action space. Regarding training stages, Stage 1 (SFT) delivers the most substantial performance leap, proving that initializing the model via expert logic distillation is essential for establishing foundational capabilities. For alignment, Stage 2 (SFT+DPO) consistently surpasses Rejection Fine-Tuning (SFT+RFT), particularly in complex reasoning tasks. We further validate the intrinsic quality of DPO-optimized meta plans through a head-to-head semantic evaluation provided in Appendix \ref{app:ablation_details}. Additionally, we verify the necessity of the dual-awareness mechanism in Appendix \ref{app:ablation_details}, where experimental results demonstrate that multi-modal inputs yield a clear performance advantage over text-only baselines. Finally, the Teacher Plan utilizes ground-truth logic, establishing the theoretical upper bound for our optimization.

\begin{table}[t]
\caption{Ablation study of the training strategy on EarthBench using Qwen3-235B. Metrics include process-level (TSA, ASF1, ACF) and outcome-level scores. \textbf{Bold} and \underline{underline} denote the best and second-best results among trained models (excluding the \textit{Teacher Plan}). Tool-RAG represents a baseline utilizing semantic retrieval without hierarchical planning.}
\label{tab:ablation_main}
\centering
\small
\setlength{\tabcolsep}{3pt}
\begin{tabular}{l|ccc|cccc}
\toprule
\textbf{Configuration} & \textbf{TSA} & \textbf{ASF1} & \textbf{ACF} & \textbf{FAA} & \textbf{TAO} & \textbf{TIO} & \textbf{TEM} \\
\midrule
Base     & 0.537 & 0.479 & 0.421 & 0.261 & 0.618 & 0.458 & 0.183 \\
Tool-RAG   & 0.567 & 0.516 & 0.445 & 0.304 &  0.646 &  0.507  & 0.166 \\
Zero-Shot      & 0.626 & 0.574 & 0.517 & 0.304 & \underline{0.682} & \underline{0.592} & 0.248 \\
\midrule
SFT           & 0.685 & 0.646 & \textbf{0.608} & 0.304 & 0.665 & 0.534 & 0.405 \\
SFT+RFT          & \underline{0.689} & \underline{0.650} & 0.585 & \textbf{0.326} & 0.662 & 0.569 & \textbf{0.437} \\
\textbf{SFT+DPO} & \textbf{0.698} & \textbf{0.655} & \underline{0.604} & \textbf{0.326} & \textbf{0.736} & \textbf{0.612} & \underline{0.429} \\
\midrule
\textit{Teacher Plan}  & \textit{0.751} & \textit{0.708} & \textit{0.652} & \textit{0.348} & \textit{0.766} & \textit{0.687} & \textit{0.498} \\
\bottomrule
\end{tabular}
\end{table}

\subsection{Scalability and Generalizability across Model Capacities}
\label{sec:scalability}

\begin{figure}[t]
    \vskip -0.1in
    \centering
    \includegraphics[width=0.9\columnwidth]{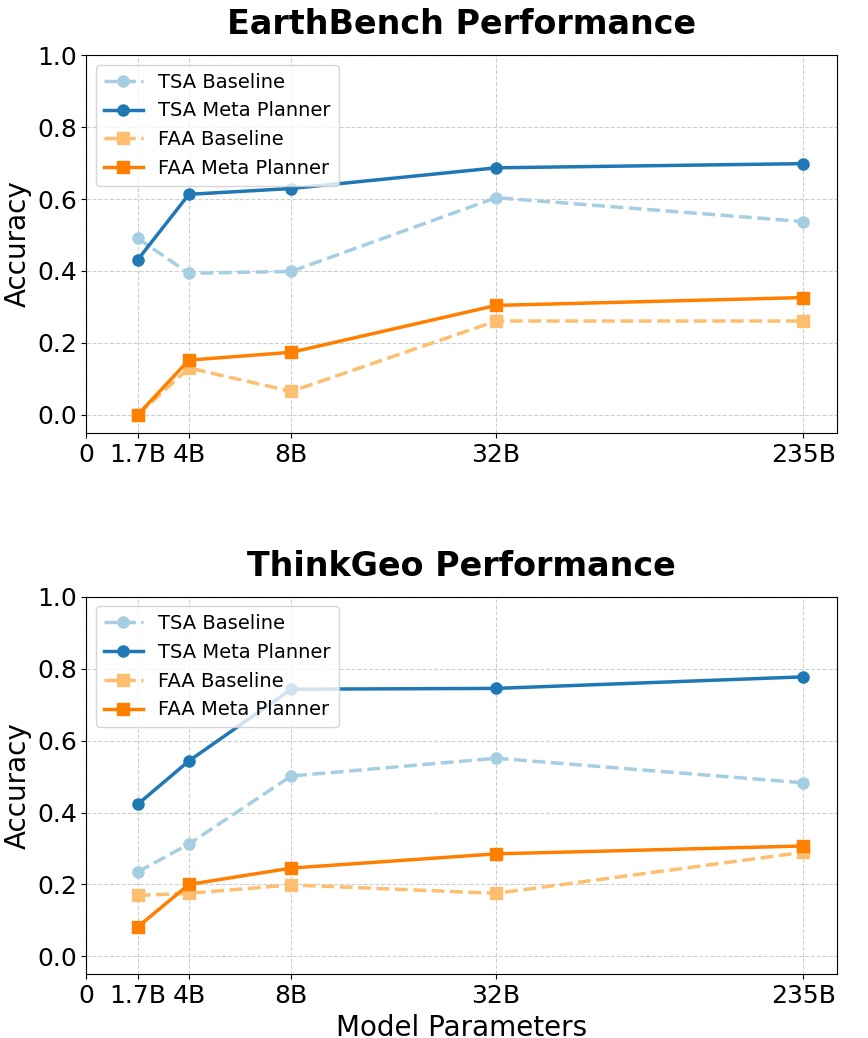}
    \caption{Scalability analysis of the Meta-Planner across Qwen3 Executor models. The charts display TSA and FAA on EarthBench (top) and ThinkGeo (bottom). The Meta-Planner (solid lines) consistently outperforms the Baseline (dashed lines) as model capacity increases.}
    \label{fig:executor_scale}
    \vskip -0.5in
\end{figure}

To validate the versatility of our framework, we conducted experiments on the Qwen3-Instruct family with parameter sizes ranging from 1.7B to 235B. Figure \ref{fig:executor_scale} illustrates the performance trends for TSA and FAA on EarthBench and ThinkGeo. Detailed numerical results are provided in Appendix \ref{app:scalability_table}.

We observe that the Meta-Planner generally enhances performance across the model spectrum, with the magnitude of improvement varying by size. For the 1.7B model, the insertion of a meta plan results in performance stagnation or slight degradation. This behavior likely stems from the limited instruction-following capabilities of extremely small models, which struggle to comprehend and execute structured expert guidance effectively. In contrast, significant gains appear in medium-sized agents such as the 4B and 8B models. For instance, the TSA of the 4B model on EarthBench improves substantially, enabling this lightweight model to rival the baseline performance of larger counterparts. Furthermore, unlike general domain tasks where benefits often diminish for the largest models, our framework continues to yield improvements for the 235B model. This suggests that domain-specific remote sensing logic is distinct from general reasoning capacity, and the injected expert constraints provide value even to massive foundation models.

\subsection{Reliability Analysis of Automated Evaluation}
\label{sec:human_align}

\begin{figure}[h]
    \vskip -0.1in
    \centering
    \includegraphics[width=0.95\columnwidth]{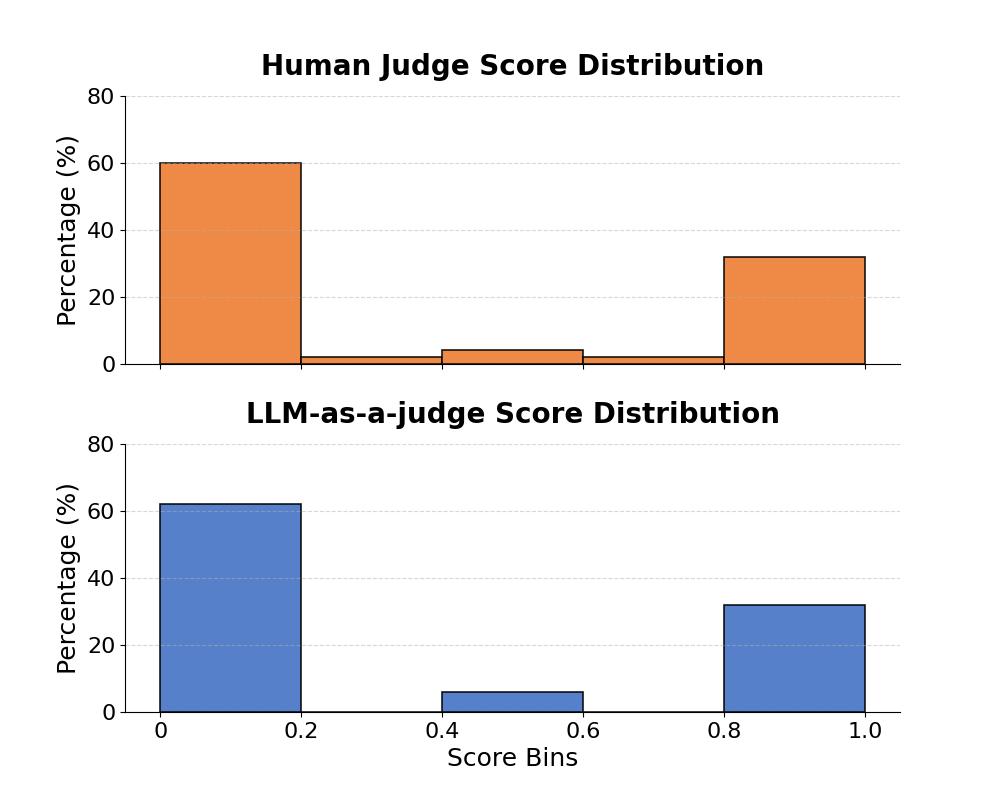}
    \caption{Alignment analysis of score distributions between Human Judges (top) and the LLM-as-a-judge (bottom) across 50 randomly sampled tasks.}
    \label{fig:judge_alignment}
    \vskip -0.5in
\end{figure}

To validate the reliability of the LLM-as-a-judge protocol, we conducted a human alignment study using 50 randomly sampled execution trajectories from the test sets. These samples were manually graded by five remote sensing domain experts. As illustrated in Figure \ref{fig:judge_alignment}, the score distribution generated by the LLM closely mirrors the human evaluation. Both distributions exhibit a distinct bimodal pattern typical of agentic tasks. The lack of significant divergence suggests that the automated judge introduces no systematic bias. This alignment confirms that our automated metrics serve as a reliable proxy for expert judgment, reinforcing the credibility of the reported results.

\section{Conclusion}
\label{sec:conclusion}

In this work, we presented a Lightweight Multimodal Meta-Planner framework to resolve the reasoning limitations and cognitive overload in EO agents that integrate planning and execution within a single model. By decoupling high-level reasoning from precise execution and injecting expert knowledge via the Meta Task Library, our approach enables the Meta-Planners to navigate complex tool spaces effectively. We implemented a two-stage training pipeline that initializes the Meta-Planner through expert logic internalization and refines it via execution-feedback-driven preference alignment. Evaluations on EarthBench, ThinkGeo, and an out-of-domain test set demonstrate that our approach significantly improves tool-calling accuracy and generalization. Furthermore, our experiments validate the ``small-drives-large'' paradigm, showing that a specialized 8B Meta-Planner consistently enhances diverse executor backbones ranging from 1.7B to 235B parameters without costly full-parameter fine-tuning.

\textbf{Limitations.} Despite achieving superior performance compared to baselines, several challenges remain. (1) Data Scarcity and Variance: The field currently relies on a restricted number of high-quality benchmarks. This scarcity not only limits the scope of training but also introduces high variance in evaluation results due to small sample sizes. (2) Gap to Industrial Deployment: While our framework significantly improves intermediate process accuracy, the End-to-End FAA still falls short of the rigorous standards required for real-world application, underscoring the need for future work on error recovery mechanisms and robust backend systems. (3) Model Efficiency Exploration: Our current Meta-Planner relies on an 8B backbone. Future research could explore the boundaries of parameter efficiency, investigating whether extremely lightweight models can retain sufficient strategic reasoning capabilities to further reduce computational costs.








\section*{Impact Statement}


This paper presents work whose goal is to advance the field of Machine
Learning. There are many potential societal consequences of our work, none
which we feel must be specifically highlighted here.


\nocite{langley00}

\bibliography{example_paper}
\bibliographystyle{icml2026}

\newpage
\appendix
\onecolumn
\section{Hyperparameter Analysis for Step-aware Discounted Reward}
\label{sec:appendix_gamma}

In the \textbf{Preference Alignment (DPO)} stage, we employ a Step-aware Discounted Reward to prioritize early-step correctness during trajectory evaluation. The discount factor $\gamma$ controls the decay rate of the reward weight over time steps. To determine the optimal $\gamma$, we conducted a sensitivity analysis by varying $\gamma$ from 0.1 to 1.0. We assessed the impact on the construction of the preference dataset $\mathcal{D}_{\text{DPO}}$ using three metrics:
\begin{enumerate}
    \item \textbf{Number of Preference Pairs (NPP):} The total number of valid preference pairs $(p_w, p_l)$ extracted under a specific $\gamma$.
    \item \textbf{Sample Overlap Rate (SOR):} The ratio of identical preference pairs shared between the current $\gamma$ setting and the baseline setting ($\gamma=0.9$).
    \item \textbf{Preference Turn Rate (PTR):} The number of pairs where the preference direction flips compared to the baseline (i.e., a plan considered ``winning'' at $\gamma=0.9$ becomes ``losing'' at the current $\gamma$).
\end{enumerate}

\begin{figure}[h]
    \centering
    \begin{minipage}{0.32\textwidth}
        \centering
        \includegraphics[width=\linewidth]{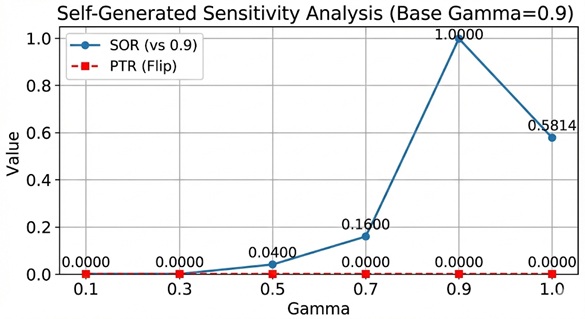}
        \caption*{Self-Generated Sensitivity}
    \end{minipage}
    \hfill
    \begin{minipage}{0.32\textwidth}
        \centering
        \includegraphics[width=\linewidth]{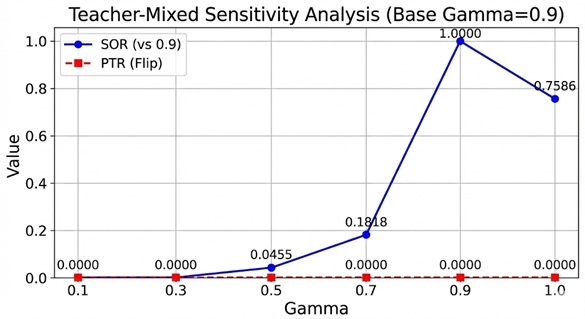}
        \caption*{Teacher-Mixed Sensitivity}
    \end{minipage}
    \hfill
    \begin{minipage}{0.32\textwidth}
        \centering
        \includegraphics[width=\linewidth]{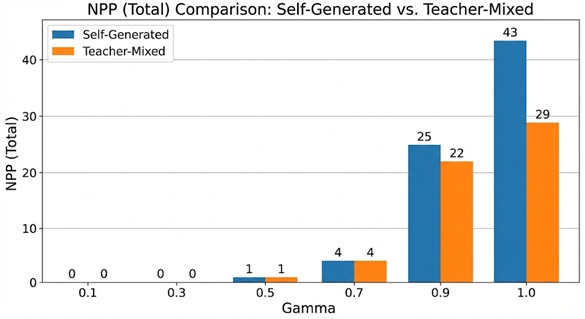}
        \caption*{NPP Comparison}
    \end{minipage}
    \caption{Hyperparameter sensitivity analysis for the discount factor $\gamma$. The plots illustrate the Sample Overlap Rate (SOR) and Preference Turn Rate (PTR) for Self-Generated (Left)  and Teacher-Mixed (Center)  datasets. The bar chart (Right)  displays the total Number of Preference Pairs (NPP) retained at each $\gamma$ level.}
    \label{fig:gamma_analysis}
\end{figure}

The results of this analysis are presented in Figure~\ref{fig:gamma_analysis}. We observe distinct behavioral trends as $\gamma$ shifts from 0.1 to 1.0.

First, the physical interpretation of $\gamma$ dictates the reward distribution. A $\gamma$ approaching 0 assigns exclusive weight to the initial steps, while a $\gamma$ of 1.0 treats all steps equally by calculating a simple average. Our objective is to prioritize the foundational logic in early steps without completely discarding the value of subsequent actions.

Second, the \textbf{NPP} metric reveals a critical constraint regarding data availability. As shown in the rightmost panel of Figure~\ref{fig:gamma_analysis}, the number of valid preference pairs exhibits a precipitous decline as $\gamma$ decreases below 0.9. At $\gamma=0.7$ and lower, the strict penalty on later steps filters out the vast majority of candidate pairs, resulting in extreme data scarcity. To ensure the DPO training pipeline has sufficient data density for effective optimization, maintaining a high NPP is essential.

Third, comparing $\gamma=0.9$ with $\gamma=1.0$ validates the necessity of the discount mechanism. The \textbf{SOR} values for both Self-Generated (0.58) and Teacher-Mixed (0.75) datasets indicate that the dataset constructed at $\gamma=0.9$ differs significantly from the non-discounted average ($\gamma=1.0$). This divergence confirms that $\gamma=0.9$ successfully filters out pairs where performance gains in later steps mask failures in early reasoning, thereby enforcing the intended temporal logic.

Finally, the \textbf{PTR} remains constant at zero across all settings. This stability indicates that the preference labels are robust. Varying the discount factor does not invert the judgment of ground truth; rather, it refines the threshold for statistical significance.

Based on these observations, we select $\gamma=0.9$ as the optimal hyperparameter. This setting effectively balances the requirement for sufficient training data (High NPP) with the need to rigorously enforce step-wise causal dependencies (Moderate SOR vs 1.0), without introducing label noise (Zero PTR).

\section{Dataset Details}
\label{app:datasets}

This section details the specifications for the three datasets used in our evaluation: EarthBench, ThinkGeo, and our custom GeoScenario-116.

\subsection{EarthBench}
\textbf{EarthBench}~\cite{feng2025earth} is a benchmark designed to evaluate the multi-step reasoning and tool usage capabilities of Earth Observation (EO) agents. Unlike traditional datasets that focus solely on perception, EarthBench integrates three data modalities: \textit{RGB Imagery}, \textit{Raw Spectral Data}, and \textit{Processed Earth Products}. The dataset employs a dual-level evaluation protocol that assesses both the correctness of the final answer and the logical validity of the reasoning trajectory.

The benchmark contains 104 expert tools organized into five functional categories. Table~\ref{tab:earthbench_tools} summarizes the tool distribution and typical tasks. Our experiments use a standard split of 200 training samples and 46 test samples. The test set includes 31 zero-shot tools unseen during training, which evaluates the generalization ability of the agent across different physical interpretation tasks.

\begin{table}[h]
    \centering
    \caption{Tool Categories and Statistics in EarthBench}
    \label{tab:earthbench_tools}
    \small 
    \begin{tabularx}{0.9\textwidth}{@{}llX@{}} 
        \toprule
        \textbf{Category} & \textbf{Count} & \textbf{Description \& Examples} \\
        \midrule
        \textbf{Index} & 12 & Tools for computing remote sensing indices such as \texttt{compute\_ndvi} and \texttt{compute\_ndwi} to characterize vegetation, water, and burn severity. \\
        \textbf{Inversion} & 18 & Algorithms for retrieving geophysical parameters from raw spectral data, including \texttt{retrieve\_lst} and \texttt{invert\_pwv}. \\
        \textbf{Perception} & 15 & Vision-oriented tools for scene classification, object detection, and segmentation like \texttt{count\_objects} and \texttt{segment\_water}. \\
        \textbf{Analysis} & 10 & Tools for spatiotemporal reasoning and time-series analysis such as \texttt{detect\_trend} and \texttt{seasonal\_decomposition}. \\
        \textbf{Statistics} & 49 & Functions for data preprocessing, statistical computation, and cloud masking, including \texttt{calculate\_mean} and \texttt{mask\_clouds}. \\
        \bottomrule
    \end{tabularx}
\end{table}

\subsection{ThinkGeo}
\textbf{ThinkGeo}~\cite{shabbir2025thinkgeo} evaluates tool-augmented agents on geospatial tasks requiring spatial reasoning and multi-step planning. It features high-resolution optical imagery and questions that demand the coordination of perception, logic, and operation tools.

We curated a subset of 387 samples from the original benchmark, consisting of 330 training and 57 testing samples, where tool execution exhibits high determinism. Table~\ref{tab:thinkgeo_scenarios} details the seven task scenarios. The agent must use a set of 14 tools, such as \texttt{ObjectDetection}, \texttt{ChangeDetection}, and \texttt{GeographyMath}, to derive open-ended conclusions. This dataset challenges the ability of the agent to interpret complex urban and environmental contexts.

\begin{table}[h]
    \centering
    \caption{Scenario Categories in ThinkGeo}
    \label{tab:thinkgeo_scenarios}
    \small 
    \begin{tabularx}{0.9\textwidth}{@{}lX@{}}
        \toprule
        \textbf{Scenario} & \textbf{Task Description} \\
        \midrule
        \textbf{Urban Planning} & Analysis of residential layouts, road mapping, and infrastructure density. \\
        \textbf{Disaster Assessment} & Damage evaluation and change detection following events like floods or hurricanes. \\
        \textbf{Environmental Monitoring} & Assessment of water bodies, vegetation health, and pollution sources. \\
        \textbf{Transportation Analysis} & Traffic flow estimation, vehicle counting, and intersection analysis. \\
        \textbf{Aviation Monitoring} & Aircraft identification, runway occupancy, and airport logistics. \\
        \textbf{Recreational Infrastructure} & Identification and coverage estimation of sports fields and parks. \\
        \textbf{Industrial Sites} & Localization of storage tanks and analysis of industrial operational zones. \\
        \bottomrule
    \end{tabularx}
\end{table}

\subsection{GeoScenario-116}
To evaluate the generalization capabilities of our agent, we constructed a custom out-of-domain test set containing 116 samples. This dataset relies on a distinct subset of 9 offline-deployable tools. The samples focus on specialized scenes, specifically Airports and Harbors. These environments differ significantly from the general urban and rural scenes dominant in the training data and serve as a testbed for agent robustness in novel environments.

\section{Additional Ablation and Evaluation Details}
\label{app:ablation_details}

This section provides supplementary experimental results verifying the impact of multi-modal inputs and a detailed head-to-head quality assessment of the training stages.

\begin{table}[t]
\setlength{\intextsep}{0pt}
\vspace{-9pt}
\caption{Ablation study on the impact of multi-modal inputs on the ThinkGeo dataset (Executor: Qwen3-8B). The integration of visual features consistently improves planning accuracy compared to text-only inputs.}
\label{tab:multimodal_ablation}
\centering
\small
\begin{tabular}{ll|cccccc}
\toprule
\textbf{Stage} & \textbf{Modality} & \textbf{TSA} & \textbf{ASF1} & \textbf{ACF} & \textbf{FAA} & \textbf{TAO} & \textbf{TEM} \\
\midrule
No-Train & Text-Only & 0.554 & 0.554 & 0.411 & \textbf{0.289} & 0.716 & 0.293 \\
         & Multi-Modal & 0.614 & 0.613 & 0.452 & 0.234 & 0.728 & 0.370 \\
\midrule
SFT+DPO  & Text-Only & \underline{0.693} & \underline{0.692} & \underline{0.498} & 0.189 & \underline{0.836} & \textbf{0.576} \\
         & \textbf{Multi-Modal} & \textbf{0.743} & \textbf{0.741} & \textbf{0.519} & \underline{0.246} & 0.769 & \underline{0.506} \\
\bottomrule
\end{tabular}
\vspace{-20pt}
\end{table}

\subsection{Impact of Multi-Modal Inputs}
To isolate the contribution of the \textbf{Dual-awareness Mechanism}, we evaluated the Meta-Planner on the ThinkGeo dataset using the Qwen3-8B executor under two distinct conditions. The \textit{text-only} setting relies on image captions provided by the benchmark, whereas the \textit{multi-modal} setting utilizes direct visual embeddings. As presented in Table \ref{tab:multimodal_ablation}, the multi-modal configuration outperforms the text-only variant in both zero-shot and DPO phases. These results validate the effectiveness of incorporating multi-modal information, evidenced by consistent gains across both process-level and outcome-level metrics.

\subsection{Head-to-Head Plan Quality Assessment}
To directly evaluate the intrinsic quality of the generated meta plans independent of downstream execution success, we employed a head-to-head comparison protocol adopted from MPO \cite{xiong2025mpo}. We used GPT-4o as a professional evaluator to compare plans generated by the SFT model against those from the SFT+DPO model.

\begin{figure}[h]
    \centering
    \includegraphics[width=0.8\columnwidth]{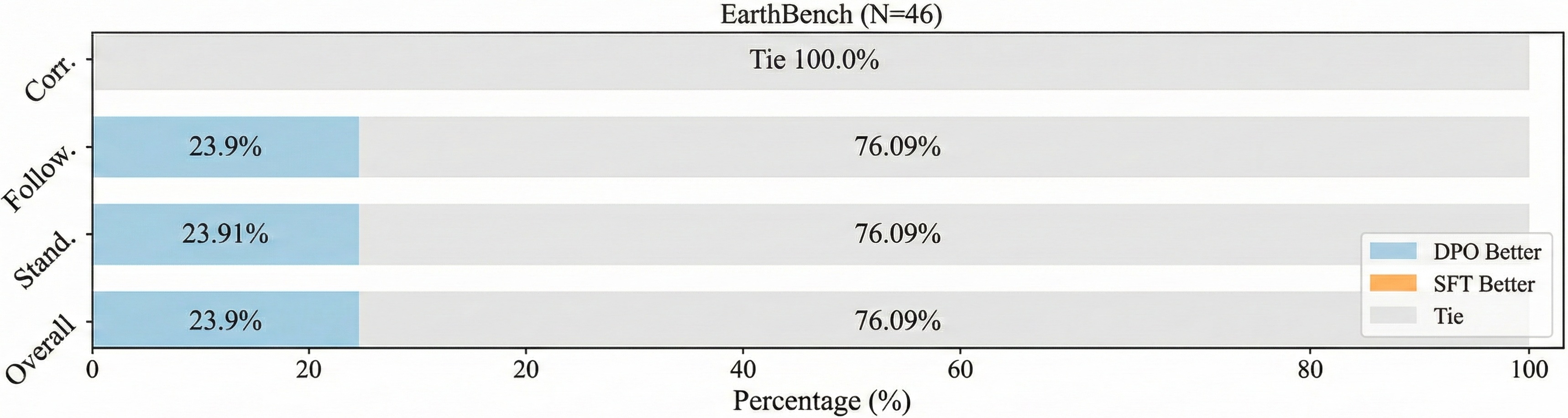}
    \caption{Head-to-head qualitative comparison of meta plan quality between SFT and DPO stages on EarthBench ($N=46$). Evaluated by GPT-4o, the DPO-optimized Meta-Planner demonstrates strict dominance, significantly improving \textbf{Followability} and \textbf{Standardization} (23.9\% win rate) while maintaining perfect parity in \textbf{Correctness} (100\% Tie), with zero regression (0\% Loss).}
    \label{fig:direct_eval}
\end{figure}

As illustrated in Figure \ref{fig:direct_eval}, the DPO-optimized Meta-Planner demonstrates a \textbf{23.9\%} strict win rate over the SFT baseline, with zero instances of regression (0\% Loss). The high tie rate (76.09\%) suggests that while SFT successfully establishes the correct basic structure, the magnitude of further improvement is partially constrained by the limited size of the preference dataset. Nevertheless, DPO acts as a crucial precision refiner, explicitly enhancing the standardization and followability of the plans in complex scenarios.

The automated assessment was conducted using the specific instruction prompt detailed below:

\begin{tcolorbox}[colback=gray!10, colframe=gray!50, title=Instruction Prompt for GPT Automated Assessment]
\small
\texttt{Please act as a professional instruction evaluator and assess the following two sets of meta plans.}

\texttt{Task description: \{task\}}

\texttt{DPO Plan: \{dpo\}}

\texttt{SFT Plan: \{sft\}}

\texttt{Please compare these two sets of meta plans across the following three dimensions:}

\texttt{1. Correctness - Does the meta plan accurately fulfill the task requirements?}

\texttt{2. Followability - Is the meta plan clear, easy to understand, and are the steps reasonable?}

\texttt{3. Standardization - Does the meta plan follow a consistent and standardized format?}

\texttt{For each dimension, please indicate which meta plan is better and provide reasoning. Finally, provide an overall assessment.}

\texttt{Please output the result in JSON format, including the following fields:}

\begin{verbatim}
{ 
    "correctness_better": "dpo"/"sft"/"tie", 
    "correctness_reason": "reason", 
    "followability_better": "dpo"/"sft"/"tie", 
    "followability_reason": "reason", 
    "standardization_better": "dpo"/"sft"/"tie", 
    "standardization_reason": "reason", 
    "overall_better": "dpo"/"sft"/"tie" 
}
\end{verbatim}
\end{tcolorbox}

\section{Detailed Scalability Results}
\label{app:scalability_table}

Table \ref{tab:scalability_detailed} provides the comprehensive metrics for the scalability analysis discussed in Section \ref{sec:scalability}, covering both process-level and outcome-level evaluations across the Qwen3 model family.

\begin{table*}[h]
\caption{Scalability analysis of the Meta-Planner across Qwen3 Executor models. Metrics evaluate \textbf{Process Metrics} (TSA, ASF1, ACF) and \textbf{Outcome Metrics} (FAA, TAO, TIO, TEM). Columns with a gray background denote our Meta-Planner approach. \textbf{Bold} indicates the best result, and \underline{underlined} indicates the second best per dataset.}
\label{tab:scalability_detailed}
\centering
\small
\setlength{\tabcolsep}{0pt}
\renewcommand{\arraystretch}{1.15}
\begin{tabular*}{\textwidth}{@{\extracolsep{\fill}} l cc cc cc c cc cc cc cc @{}}
\toprule
\multirow{2}{*}{\textbf{Model}} & \multicolumn{2}{c}{\textbf{TSA}} & \multicolumn{2}{c}{\textbf{ASF1}} & \multicolumn{2}{c}{\textbf{ACF}} & & \multicolumn{2}{c}{\textbf{FAA}} & \multicolumn{2}{c}{\textbf{TAO}} & \multicolumn{2}{c}{\textbf{TIO}} & \multicolumn{2}{c}{\textbf{TEM}} \\
\cmidrule{2-7} \cmidrule{9-16}
 & Base & \cellcolor{rowgray}Meta & Base & \cellcolor{rowgray}Meta & Base & \cellcolor{rowgray}Meta && Base & \cellcolor{rowgray}Meta & Base & \cellcolor{rowgray}Meta & Base & \cellcolor{rowgray}Meta & Base & \cellcolor{rowgray}Meta \\
\midrule
\multicolumn{16}{l}{\textit{\textbf{EarthBench} (104 Tools)}} \\
\midrule
Qwen3-235B & 0.537 & \cellcolor{rowgray}\textbf{0.698}\up & 0.479 & \cellcolor{rowgray}\underline{0.655}\up & 0.421 & \cellcolor{rowgray}\textbf{0.604}\up && 0.261 & \cellcolor{rowgray}\textbf{0.326}\up & 0.618 & \cellcolor{rowgray}\textbf{0.736}\up & 0.458 & \cellcolor{rowgray}\textbf{0.612}\up & 0.183 & \cellcolor{rowgray}\textbf{0.429}\up \\
Qwen3-32B  & 0.604 & \cellcolor{rowgray}\underline{0.687}\up & 0.572 & \cellcolor{rowgray}\textbf{0.673}\up & \underline{0.585} & \cellcolor{rowgray}\underline{0.585}~ && 0.261 & \cellcolor{rowgray}\underline{0.304}\up & \underline{0.656} & \cellcolor{rowgray}0.620\down & \underline{0.534} & \cellcolor{rowgray}0.499\down & 0.290 & \cellcolor{rowgray}\underline{0.362}\up \\
Qwen3-8B   & 0.399 & \cellcolor{rowgray}0.629\up & 0.369 & \cellcolor{rowgray}0.592\up & 0.279 & \cellcolor{rowgray}0.530\up && 0.065 & \cellcolor{rowgray}0.174\up & 0.448 & \cellcolor{rowgray}0.531\up & 0.358 & \cellcolor{rowgray}0.440\up & 0.034 & \cellcolor{rowgray}0.261\up \\
Qwen3-4B   & 0.394 & \cellcolor{rowgray}0.613\up & 0.361 & \cellcolor{rowgray}0.579\up & 0.276 & \cellcolor{rowgray}0.525\up && 0.130 & \cellcolor{rowgray}0.152\up & 0.533 & \cellcolor{rowgray}0.498\down & 0.385 & \cellcolor{rowgray}0.401\up & 0.027 & \cellcolor{rowgray}0.324\up \\
Qwen3-1.7B & 0.491 & \cellcolor{rowgray}0.431\down & 0.479 & \cellcolor{rowgray}0.428\down & 0.444 & \cellcolor{rowgray}0.408\down && 0.000 & \cellcolor{rowgray}0.000~ & 0.306 & \cellcolor{rowgray}0.275\down & 0.261 & \cellcolor{rowgray}0.237\down & 0.240 & \cellcolor{rowgray}0.237\down \\
\midrule
\multicolumn{16}{l}{\textit{\textbf{ThinkGeo} (14 Tools)}} \\
\midrule
Qwen3-235B & 0.483 & \cellcolor{rowgray}\textbf{0.777}\up & 0.455 & \cellcolor{rowgray}\underline{0.743}\up & 0.392 & \cellcolor{rowgray}\textbf{0.550}\up && 0.290 & \cellcolor{rowgray}\textbf{0.307}\up & \underline{0.845} & \cellcolor{rowgray}\textbf{0.886}\up & \underline{0.691} & \cellcolor{rowgray}\textbf{0.767}\up & 0.348 & \cellcolor{rowgray}\textbf{0.625}\up \\
Qwen3-32B  & 0.551 & \cellcolor{rowgray}\underline{0.745}\up & 0.549 & \cellcolor{rowgray}\textbf{0.744}\up & 0.357 & \cellcolor{rowgray}0.488\up && 0.175 & \cellcolor{rowgray}\underline{0.285}\up & 0.684 & \cellcolor{rowgray}0.794\up & 0.548 & \cellcolor{rowgray}0.688\up & 0.238 & \cellcolor{rowgray}\underline{0.528}\up \\
Qwen3-8B   & 0.502 & \cellcolor{rowgray}0.743\up & 0.501 & \cellcolor{rowgray}0.741\up & 0.348 & \cellcolor{rowgray}\underline{0.519}\up && 0.199 & \cellcolor{rowgray}0.246\up & 0.769 & \cellcolor{rowgray}0.769~ & 0.606 & \cellcolor{rowgray}0.659\up & 0.148 & \cellcolor{rowgray}0.506\up \\
Qwen3-4B   & 0.313 & \cellcolor{rowgray}0.543\up & 0.313 & \cellcolor{rowgray}0.543\up & 0.234 & \cellcolor{rowgray}0.390\up && 0.175 & \cellcolor{rowgray}0.200\up & 0.481 & \cellcolor{rowgray}0.583\up & 0.389 & \cellcolor{rowgray}0.469\up & 0.029 & \cellcolor{rowgray}0.218\up \\
Qwen3-1.7B & 0.235 & \cellcolor{rowgray}0.423\up & 0.231 & \cellcolor{rowgray}0.421\up & 0.148 & \cellcolor{rowgray}0.241\up && 0.170 & \cellcolor{rowgray}0.082\down & 0.456 & \cellcolor{rowgray}0.415\down & 0.298 & \cellcolor{rowgray}0.253\down & 0.067 & \cellcolor{rowgray}0.056\down \\
\bottomrule
\end{tabular*}
\begin{flushleft}
\scriptsize \textbf{Note:} \up (red) denotes improvement, \down (green) denotes decline compared to baseline. Best values are \textbf{bolded} and second-best are \underline{underlined} within each dataset group.
\end{flushleft}
\end{table*}

In our scalability experiments (Table \ref{tab:scalability_detailed}), we observed a counter-intuitive phenomenon: the Qwen3-235B Baseline consistently underperforms the smaller Qwen3-32B Baseline in Tool Selection Accuracy (TSA) across both EarthBench and ThinkGeo datasets. However, when equipped with the LMMP framework, the 235B model reclaims its expected performance lead.

Our preliminary analysis suggests this stems from a \textbf{``Semantic Over-association''} effect. The 235B model, possessing stronger capabilities in context comprehension, tends to correlate high-level task keywords in the user query with sophisticated concepts in the tool descriptions, even when a simpler tool is logically required. In contrast, the 32B model, having a more constrained attention scope, is more likely to adhere to the immediate, literal instruction.

We illustrate this with a representative case from the ThinkGeo dataset (Flood Scenario), formatted using the ms-swift protocol \cite{zhao2025swift}.

\begin{tcolorbox}[colback=gray!5, colframe=gray!40, title=Case Study: Semantic Over-association in Qwen3-235B]
\small
\textbf{Context:} The agent is asked: \textit{``Are the two cars parked outside flooded house, also flooded?''}
\\
\textbf{Current Step:} Check the second car for flooding status.
\\
\textbf{Ground Truth Tool:} \texttt{RegionAttributeDescription} (Simple attribute check).
\\
\textbf{Qwen3-32B Action (Correct):} 
\begin{verbatim}
{"name": "RegionAttributeDescription", 
 "arguments": {"image": "...", "bbox": "...", "attribute": "flooded"}}
\end{verbatim}
\textit{Reasoning:} The model correctly identifies the need to check a specific attribute of a region.
\\
\textbf{Qwen3-235B Action (Incorrect):}
\begin{verbatim}
{"name": "ChangeDetection", 
 "arguments": {"pre_image": "...", "text": "Check if cars... are flooded", 
 "post_image": "..."}}
\end{verbatim}
\textit{Error Analysis:} The tool description for \texttt{ChangeDetection} contains high-level keywords such as \textit{``disaster type classification''} and \textit{``damage-severity assessment.''} The 235B model over-associates the concept of ``flooded house'' with these disaster-related keywords, hallucinating a need for multi-temporal change detection even though the task only requires static attribute recognition on a single image.
\end{tcolorbox}

The integration of the LMMP framework rectifies this by injecting expert-defined operational constraints. The Meta-Planner explicitly decomposes the task into a \texttt{Detection} $\rightarrow$ \texttt{Attribute} sequence, forcing the 235B executor to adhere to the strict domain logic rather than relying on loose semantic matching.

Interestingly, this degradation disappears in End-to-End metrics (FAA), where the 235B Baseline outperforms the 32B Baseline. We hypothesize that the step-by-step evaluation (TSA) enforces a strict adherence to human-annotated ``Golden Paths.'' In the open-loop End-to-End setting, the 235B model utilizes its superior reasoning to find alternative valid paths or self-correct during subsequent steps, thereby bypassing the specific rigid penalties incurred during teacher-forcing evaluation. We note that these observations are based on preliminary case studies, and further investigations are underway to rigorously quantify this behavior.

\end{document}